# Cyber Risks to Next-Gen Brain-Computer Interfaces: Analysis and Recommendations

Tyler Schroder[1,2], Renee Sirbu[2], Sohee Park[1], Jessica Morley[2], Sam Street[3], Luciano Floridi[2,4]

[1] Department of Computer Science, Yale University, 51 Prospect St, New Haven, CT 06511

[2] Digital Ethics Center, Yale University, 85 Trumbull St, New Haven, CT 06511

[3] Program in the History of Science and Medicine, Section of the History of Medicine, Yale University, P.O. Box 208015, New Haven, CT 06520-8015

[4] Department of Legal Studies, University of Bologna, Via Zamboni 27/29, 40126 Bologna, Italy

Email for correspondence: tyler.schroder@yale.edu

## Abstract

Brain-computer interfaces (BCIs) show enormous potential for advancing personalized medicine. However, BCIs also introduce new avenues for cyber-attacks or security compromises. In this article, we analyze the problem and make recommendations for device manufacturers to better secure devices and to help regulators understand where more guidance is needed to protect patient safety and data confidentiality. Device manufacturers should implement the prior suggestions in their BCI products. These recommendations help protect BCI users from undue risks, including compromised personal health and genetic information, unintended BCI-mediated movement, and many other cybersecurity breaches. Regulators should mandate non-surgical device update methods, strong authentication and authorization schemes for BCI software modifications, encryption of data moving to and from the brain, and minimize network connectivity where possible. We also design a hypothetical, average-case threat model that identifies possible cybersecurity threats to BCI patients and predicts the likeliness of risk for each category of threat. BCIs are at less risk of physical compromise or attack but are vulnerable to remote attack; we focus on possible threats via network paths to BCIs and suggest technical controls to limit network connections.

## Keywords

## 1. Introduction

Brain-computer interfaces (BCIs) are medical devices that can record from and stimulate the brain (Shupe et al., 2021a). They offer many therapeutic benefits (Wolpaw & Wolpaw, 2012), including reversal of seizure onset (Karageorgos et al., 2020a), enhanced motor control (i.e., for Parkinson's patients) (Lebedev & Nicolelis, 2006), and deep-brain stimulation for treatment-resistant mental illnesses (Karageorgos et al., 2020a). In the past, BCIs were single-purpose, application-specific integrated circuits (ASICs, see Figure 1 - Flexibility vs Performance in BCI) that could only perform a single health function in a loop (Lebedev & Nicolelis, 2006; Sriram, Karageorgos, et al., 2023). More recent BCIs are approaching parity with personal computing devices to allow software modifications after implantation. BCI patients can now expect features like software updates (reprogramming), local data storage, and real-time data transmission.

BCIs have been regulated for some time. For example, in the US, they are considered Class III implantable medical devices (the highest risk tier) due to their life-supporting and sustaining nature. Class III regulations are the most stringent, but are struggling to keep up with technological change, especially with modern BCIs being networked devices with remote update and access capabilities. This means that BCIs exist in a liminal space: they are treated as implants, with tight hardware restrictions but loosely regulated onboard software. Hardware power budgets are under tighter regulatory constraints than software counterparts because the regulatory landscape has precedent in the oversight of physical implants, while more advanced software components are comparatively new.

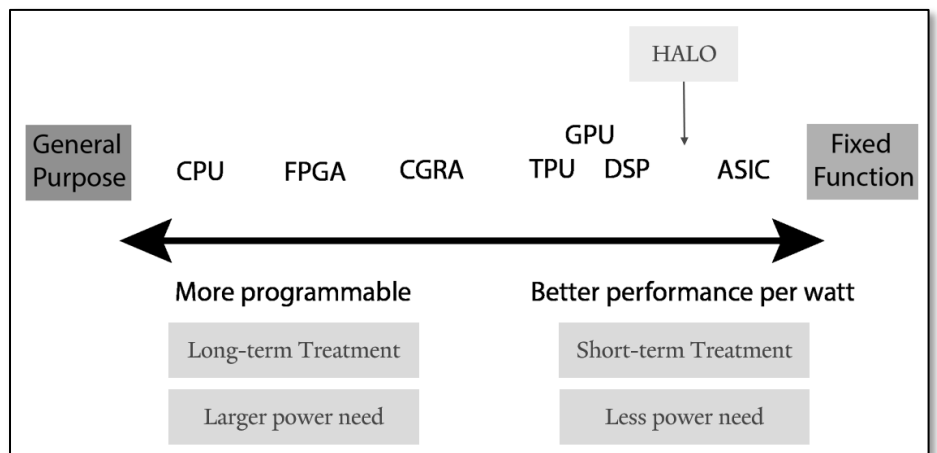

*Figure 1 - Flexibility vs Performance in BCIs*

BCIs' additional capabilities and features bring new risks. They can connect to external devices, opening them up to wireless or network attacks, and generate significant amounts of data that must be stored and protected. Other new risks concern patient safety (Sriram, Pothukuchi, et al., 2023), mental privacy (Ienca et al., 2022; McGee, 2014), autonomy (Liv, 2021), and security (Denning et al., 2009; Glannon, 2016). We address these risks elsewhere (Sirbu et al. forthcoming, 2024). In this article, we focus specifically on the cybersecurity risks posed by BCIs. We assess whether these are substantially different from the risks raised by other Class III implementable medical devices and then make some recommendations to improve the level of security for these devices.[1]

The article is structured as follows. In section two, we introduce our methodology. In section three, we outline the current landscape for Class III medical devices and general Class III medical device concerns using the US as a significant case; we then describe problems posed by BCIs and how different major regulators approach BCIs. In section four, we address broader concerns for BCI security and the neuroethical implications if security is ignored. In section five, we design a threat model for an average-case BCI and the key threats that BCIs will face. In section six, we analyze our four key problem areas from the threat model: software updates; authentication, and authorization for wireless connections; minimizing wireless attack surfaces; and encryption. We recommend actions that medical device manufacturers and regulators can take to reduce the identified risks. In section seven, we conclude the article.

---

[1] At present, BCIs are regulated as implantable medical devices instead of Software as a Medical Device (SaMD) because BCIs collect signals or data (unlike non-device software) (Health, 2023b)

## 2. Methodology

Our analysis identifies the key risks to networked BCIs by creating a threat model to understand possible attacks and their likeliness. Our model is based on a typical patient with a modern BCI implant with wireless connection capabilities. We structure our model around the four attack vectors in the *Common Vulnerability Scoring System (CVSS),* an international standard for rating vulnerabilities and security risks (FIRST, 2023).

CVSS groups attack vectors into four access types: *physical, local* (direct connection to the device), *local adjacent* (remotely from the same local network), and *network* (performed remotely anywhere in the world). Medical devices must be protected from all four vectors to minimize risk to patients. For our purposes, we will not investigate physical access threats as deeply because these devices are traditionally inaccessible.[2] Examples of local access include direct access from a keyboard/terminal, a mobile device paired with the BCI, or via a secure shell connection (a form of remote access). Local adjacent access covers access from a different machine on the same network as the BCI. Lastly, network access is any connection from devices beyond the immediate local network, such as a threat actor operating in a different network in the same building, like a coffee shop on the ground floor, or further away, like operating in a different town, province, or country.

## 3. General Cyber Risks to Class III Devices

Regarding regulation, BCIs fit into the larger category of medical devices. Using the US as a good example, the Food and Drug Administration (FDA) defines medical devices as "an instrument, apparatus, implement, machine, contrivance, implant, in vitro reagent, or other similar or related article, including a component part… intended for use in the diagnosis of disease or other conditions, or in the cure, mitigation, treatment, or prevention of disease, in man".[3] The exact method of regulation for medical devices depends on their risk classification, from low risk (Class I) to high risk (Class III). Class III devices are typically life-sustaining or supporting and have the strictest requirements and standard of evidence requirements (compared to Class I or II devices) (Zettler & Lietzan, 2021). Some examples of Class III devices include pacemakers, breast implants, cochlear implants, and BCIs. These devices are built with an average product lifetime of 10 to 30 years (FBI, 2022). Until recently, cybersecurity was rarely considered for any class of devices. Older devices (termed "legacy devices" (Chase et al., 2023)) often have no method to upgrade software or patch bugs, compared to newer devices that may support frequent updates. As most older devices ran in isolated, offline environments, this was not a primary concern for the FDA. However, this approach has changed as devices have developed network capabilities, and cyber incidents across the healthcare sector have become routine (Seh et al., 2020), this approach has changed. Beyond traditional ransomware and attacks on healthcare infrastructure,[4] the FDA has begun to warn patients and manufacturers of vulnerabilities in device communication protocols such as URGENT/11.[5] These vulnerabilities led to the October 2014 cybersecurity planning requirement: "Content of Premarket Submissions for Management of Cybersecurity in Medical Devices" (Sirbu, 2023). Up until 2022, these guidelines were only non-binding recommendations. The *Food and Drug Omnibus Reform Act of 2022 (FDORA)* and *Protecting and Transforming Cyber Health Care Act of 2022* (PATCH) introduced the enforcement by requiring cybersecurity information to be submitted alongside applications for marketing new devices to the public. (*H.R.2617 - 117th Congress (2021-2022): Consolidated Appropriations Act, 2023*, 2021; Schwartz, 2024).[6] Before the enforcement of these guidelines, many devices without appropriate cybersecurity protocols entered the market, lacking essential security functions. These devices did not employ encryption—a requirement for clinical data under the Health Insurance Portability and Protection Act (see Table 1 - HALO Device Survey (Karageorgos et al., 2020a)). Furthermore, some devices may not have had authentication and authorization schemes to ensure that only approved access to the device is possible. Without appropriate authentication, these devices were at risk of unauthorized or unwanted modification by malicious cyber actors. Now is an opportune time to suggest changes to medical device cybersecurity regulation, as BCIs have yet to hit general availability.

---

[2] Select older devices have physical SD card ports (Shupe et al., 2021b); others may have on-brain storage drives for collected data (Sriram, Pothukuchi, et al., 2023), but this will be addressed separately in the encryption recommendation.

[3] Section 201(h) of the Food, Drug, and Cosmetic Act (Health, 2023a).

---

[4] This can cause up to $900,000 of losses per day and increase strain on healthcare networks (*US Healthcare at Risk*, n.d.)

[5] (*URGENT/11 Cybersecurity Vulnerabilities in a Widely-Used Third-Party Software Component May Introduce Risks During Use of Certain Medical Devices: FDA Safety Communication*, 2019) and SweynTooth (FDA, 2020). URGENT/11 was 11 vulnerabilities to allow remote takeover of embedded medical devices. SwyenTooth was a Bluetooth connection vulnerability.

[6] Information included post market vulnerability/exploit control plans, communications to impacted patients, cybersecurity design information, and a software bill of materials (SBOM) containing all software components used in development and active use of the device

| | Medtronic [10] | Neuropace [106] | Aziz [23] | Chen [37] | Kassiri [56] | Neuralink [74] | NURIP [84] | HALO |
|---|---|---|---|---|---|---|---|---|
| **Tasks Supported** | | | | | | | | |
| Spike Detection | × | × | × | × | × | × | × | ✓ |
| Compression | × | × | ✓ | × | × | × | × | ✓ |
| Seizure Prediction | × | ✓ | × | ✓ | ✓ | × | ✓ | ✓ |
| Movement Intent | ✓ | × | × | × | × | × | × | ✓ |
| Encryption | × | × | × | × | × | × | × | ✓ |
| **Technical Capabilities** | | | | | | | | |
| Programmable | ✓ | Limited | × | Limited | ✓ | × | Limited | ✓ |
| Read Channels | 4 | 8 | 256 | 4 | 24 | 3072 | 32 | 96 |
| Stimulation Channels | 4 | 8 | 0 | 0 | 24 | 0 | 32 | 16 |
| Sample Frequency (Hz) | 250 | 250 | 5K | 200 | 7.2K | 18.6K | 256 | 30K |
| Sample Resolution (bits) | 10 | 10 | 8 | 10 | - | 10 | 16 | 16 |
| Safety (<15mW) | ✓ | ✓ | ✓ | × | ✓ | × | ✓ | ✓ |

*Table 1 - HALO Device Survey (Karageorgos et al., 2020a)*

The US is not the only regulator of BCI devices. A useful comparator comes from the European Union which regulates BCIs under both consumer protection law (General Product Safety Regulation) and medical device regulation (MDR)(Steindl, 2024).[7] Medical-purpose BCIs are rated as a EU Class III (highest risk), but are downgraded to Class II if they only "monitor physiological processes' with no ability to stimulate or influence neuroactivity (Regulation (EU) 2017/745 of the European Parliament and of the Council of 5 April 2017 on Medical Devices, Amending Directive 2001/83/EC, Regulation (EC) No 178/2002 and Regulation (EC) No 1223/2009 and Repealing Council Directives 90/385/EEC and 93/42/EEC (Text with EEA Relevance. ), 2017; Steindl, 2024). EU Class III devices require similar pre-market notification and safety requirements as the US FDA requires for US Class III devices.

These concerns are not purely theoretical, and case studies have already been analyzed in the literature. Researchers have demonstrated it is possible to attack BCI devices via adversarial input and discussed the possibility of cyber-attacks (López Madejska et al., 2024; QianQian Li et al., 2015; Upadhayay &

---

[7] Article 2(1) MDR lists specific medical purposes that require BCI devices to be regulated as medical devices instead of consumer products

Behzadan, 2023). More broadly, the healthcare sector has been hit hard by ransomware attacks, with the average cost of a data breach reaching $15 million per incident (Seh et al., 2020). A notable example of a cyber-attack on a Class III medical device was by Li (2011), who demonstrated the first remote and on-system attacks on an insulin pump (Li et al., 2011). These threats will likely increase in frequency and severity as remote monitoring and network connectivity become more common in BCIs (and medical devices more broadly) (Pycroft et al., 2016). Research has suggested that, as devices become inter-connected, moving laterally between compromised implants in a patient could be possible, allowing a threat actor to take over other implants or gain total control over someone (Ienca & Haselager, 2016).[8] Successful attacks could result in significant unwanted disclosure of personal health information, genetic data, or risks to patient safety (Liv & Greenbaum, 2023).

AI presents exciting opportunities and further risks for BCIs. Modern BCIs rely on machine learning and neural networks to decode and process signals faster and more efficiently. One such attack vector is adversarial input: intentional changes to signals designed to fool an algorithm that processes it (Finlayson et al., 2019). BCIs are shown to be vulnerable to this class of attack as "adversarial stimuli", the ability to induce changes to an environment to cause undesired output to BCIs (Upadhayay & Behzadan, 2023). These attacks do not require direct access to the patient and can be orchestrated remotely by an attacker (Upadhayay & Behzadan, 2023). The use of AI systems in BCIs requires additional care and consideration. We address recommendations for these scenarios later in the article.

## 4. Broader Risks to BCI Devices

Novel BCI devices present new challenges to neuroethics and neurorights. We discuss this further in a forthcoming article (Sirbu et al. forthcoming, *Regulating Next-Generation Implantable Brain-Computer Interfaces: Recommendations for Ethical Development and Implementation*) but highlight some key issues regarding cybersecurity concerns here. One such issue is around mental privacy. Modern BCIs are increasingly turning to wireless transmission of recorded and collected data. Improper security will put patients' mental privacy at risk. Malicious intrusion into BCI devices by attackers would result in compromised neurosecurity of the patient (Sirbu et al. forthcoming, *Regulating Next-Generation*

---

[8] In information security, lateral movement is defined as the process by which attackers spread from an entry point to the rest of the network. A threat actor gains access to one implant, and then spreads to additional within the same person (Cloudflare, n.d.).

*Implantable Brain-Computer Interfaces: Recommendations for Ethical Development and Implementation*). A significant concern for compromised neurosecurity is the ability of an attacker to induce unwanted action by a patient. Lastly, although BCIs are not currently being used to augment normal brain function, we would expect interest from various parties for BCIs being used to push abilities past neurological "normal thresholds". Secure BCI devices will prevent both biohacking and BCIs from providing "off-label" benefits to the limited population with such devices.

These concerns scale as we move up from the individual level to the country or societal level. Successful cyber-attacks target flaws in widely used programs to inflict maximum damage with minimal effort. BCIs become a valuable target for commercialization. A widespread security breach in standardized BCI systems would affect millions of users simultaneously, leading to unprecedented mass manipulation of neural data or impairment of cognitive functions. Such an attack could paralyze critical infrastructure by incapacitating key personnel, disrupt social order through mass disorientation, or even enable hostile actors to harvest sensitive thoughts and memories across an entire population. The prospect of a society-wide BCI hack represents one of the most severe cybersecurity threats imaginable, combining the scale of traditional digital attacks with direct access to human consciousness. Getting security right on these devices, before mass commercialization, remains the best defense against the doomsday laid out prior.

In this article, we have not discussed military uses for BCIs. However, this is a point of significant interest for international governing bodies. In 2020, the RAND Corporation released a 45-page initial assessment document outlining potential US military applications for BCIs and their downstream implications (Binnendijk et al., 2020). The US Defense Advanced Research Projects Agency has investigated similar applications (*Next-Generation Nonsurgical Neurotechnology*, n.d.). These applications raise independent questions on traditional and neuroethics not discussed in this article and will be the subject of a separate investigation in a future work (Sirbu et al. forthcoming, *Regulating Next-Generation Implantable Brain-Computer Interfaces: Recommendations for Ethical Development and Implementation*). Some (non-exhaustive) concerns surround the autonomy of warfighters, self-ownership and agency, the implications of emerging technologies on Just War Theory (JWT) and through International Humanitarian Law (IHL), and the complication of existing international alliances with disparate technological capabilities. These concerns diverge from the investigation of medically relevant differences between existing IMDs and BCIs and are, therefore, not addressed in our analysis.

## 5. Threat Model Analysis

As discussed, BCIs are not unique in having physiological takeover risks; insulin pumps can be remotely taken over to deliver lethal overdoses, and pacemakers could fail to stimulate the heart (Newman, 2018). We summarize these risks in Table 2 - Consequences of BCI risks and then further describe how to categorize and think about said risks.

| Consequence | Description | Unique to BCIs? | Risk Amplified by AI |
|---|---|---|---|
| Data theft or loss | Data is stolen, corrupted, or made unusable on the device. | No | Yes |
| Device disabled or unusable | Most medical devices are external to the body or easily reset. BCIs require surgical intervention if they are disabled or require a reset. | Partially | Yes |
| Physiological Takeover | A medical device performs an unwanted action under remote command | No | Yes |
| Unwanted movement | A medical device induces unwanted movement in a patient | Yes | Yes |
| Undesired brain stimulation | A medical device incorrectly applies an electric stimulation to the brain | Yes | Yes |
| Incorrect vision processing | A medical device incorrectly processes vision input (including or removing images) | Yes | Yes |
| Incorrect speech synthesis | A medical device incorrectly synthesizes speech (omitting or including words) | Yes | Yes |

Table 2 - Consequences of BCI risks

BCIs are a treasure trove of information, generating significant amounts of data they need to store. Storage options range from local storage on the implant to off-brain storage, such as a separate computer or paired companion app (Sriram, Pothukuchi, et al., 2023). There is a design trade-off when deciding on local or remote storage: local storage is faster but more expensive, while remote storage can hold more data, but takes longer to access. Research teams have previously opted for local storage to meet the access time requirement for neuron spike detection in their recent BCI design in the so-called Scalable Architecture for Low Power Devices (SCALO) (Sriram, Pothukuchi, et al., 2023). Regardless of storage location, clinical data should be encrypted to meet the HIPAA Privacy Rule (*Health Insurance Portability and Accountability Act of 1996 (HIPAA) | CDC*, 2022), which requires providers to "ensure the confidentiality, integrity, and availability of all e-PHI [electronic patient health information]" (*Health Insurance Portability and Accountability Act of 1996 (HIPAA) | CDC*, 2022). Encryption provides benefits beyond privacy: it can ensure data integrity so an attacker cannot maliciously modify stored health data (wirelessly, as shown in Figure 2 - example intrusion risks to BCI patients) (Wilton, 2021).

*Figure 2 - example intrusion risks to BCI patients*

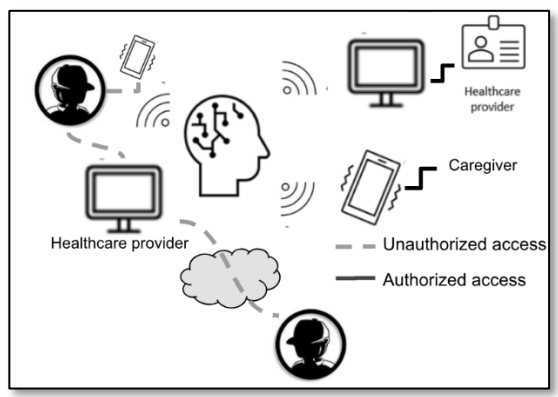

Modified data could cause patient harm by causing the BCI to take an unnecessary corrective action, such as shocking a neuron in the brain or inducing unwanted movement, like in the examples provided by Newman (2018; 2019). Furthermore, encryption also safeguards data availability. Standard encryption routines ensure that data is readily available and can be decrypted on time (Lake, 2022). A custom encryption program could take too long to access or have unknown flaws, leading to data loss.

Sriram's earlier BCI model study, Hardware Architecture for LOw-power BCIs (HALO), showed that seven of the leading BCI products on the market did not support data encryption (Karageorgos et al., 2020b) at rest (data in storage on the implant or in remote storage) or in transit (data moving between components within the BCI). Newer BCIs, like Sriram's SCALO model, allow clinicians to access real-time data and adjust the BCI programming in response (Sriram, Pothukuchi, et al., 2023). This provides enormous benefits for personalized patient care, but also opens the door to malicious exploitation, such as misuse of clinician access to send unfiltered commands to BCIs (Harris, 2020), although it presents ethical concerns about real-time remote monitoring (McGee, 2014). Proper login authentication and authorization systems can help reduce these risks.

As discussed in Section 2, we use the CVSS framework to approach risks and categorize four key areas based on these levels of access, as described below. First, local access threats affect the BCI's onboard software. BCIs may be paired to a companion device like a patient's mobile phone to support day-to-day maintenance. The companion device's core function traditionally includes software updates and real-time monitoring. Software updates are apparently simple but have risks from release to installation. Upon download to the partner device, they must be verified and transferred to the BCI to ensure the update has not been modified in transit (integrity checks).

Second, BCIs face additional threats from local adjacent network connections. Beyond the risk of loaded software, BCIs can connect to a device away from the brain to support patient care. This can involve transmitting sensor data and/or BCI configuration information (Sriram, Pothukuchi, et al., 2023). Furthermore, BCIs can also be adjusted by clinicians via local connection to better respond to patient care needs. A typical case is in seizure control when a patient no longer responds effectively to neurostimulation control (Sriram, Pothukuchi, et al., 2023). BCIs, especially older models, may not have the infrastructure in place to authenticate and authorize changes on a per-user basis. Older medical devices will assume any access is authorized. A malicious user (or even a patient) could also cause (unwanted) changes if their access is not scoped properly. This risk is also present in companion devices that a patient or clinician may use to manage the BCI. [9]

Third, next-generation BCIs like SCALO operate with multiple implants via wireless connection (Sriram, Pothukuchi, et al., 2023). Intra-BCI connections must be authenticated to ensure that each node executes only desired commands. An attacker could spoof a control signal from a primary node to other nodes in the BCI to cause unwanted movement or seizure control.

Last, BCIs face the same threats from remote attackers for network attacks. The only change is where the attacker is located. Connecting BCIs to the wider internet provides intelligent data sharing and control but at the risk of exposing the device to remote attackers. A summary of the threat model is provided in **Error! Reference source not found.**. Note that our threat model primarily focuses on average patients who are likely to be targeted by low-skill opportunistic attackers, while skilled adversaries may focus on stealing personal data and other malicious elements[10]. A high-profile individual will face more targeted attacks.

---

[9] Such as a cell phone, other mobile device, or laptop

| Attack Vector Category | Risk | Description | Unique to BCIs? |
|---|---|---|---|
| Physical | Local Storage – unencrypted data | Data is improperly accessed | No |
| Local | Software update – invalid update | An invalid software update is uploaded with the intent to disable the device | Yes |
| Local | Software update – malicious update | A malicious software update is uploaded with the intent to take over the machine | Yes |
| Local, Local Adjacent, Network | Data transmission | Data is intercepted in transit between the device and a remote computer | No |
| Local, Local Adjacent, Network | Onboard settings update | Settings are changed without authorization from the patient or clinician | Yes |
| Local, Local Adjacent, Network | Unauthorized access | Data or settings are accessed by an unwanted party | No |
| Local Adjacent, Network | Device compromise or exposure | The device is unnecessarily exposed to wireless attacks | No |

Table 3 - Summary of risks to BCI devices, by attack vector

## 6. Recommendations to Address Threat Model

BCI manufacturers and regulators can address these risks to BCI technologies *before* BCIs are generally available, and consequently harder to regulate. Our recommendations for manufacturers and regulators follow. The FIRST CVSS framework provides guidance for prioritization. If a BCI can connect to the internet or other network, the local adjacent and network risks should be addressed first. If a BCI only has local communication capability, those concerns should be addressed first. Patient-centric design should be considered as these recommendations should not limit existing healthcare functions or adoption of BCI devices. Our recommendations are summarized at the end of the section in

---

[10] See *Table 1* in (Thomas, 2022), tiers I through III. Nation-state backed actors with organized and highly technical personnel may be able to develop new threats not captured by this model.

.

## 6.1 Software Updates

Software updates must have integrity checks, a wireless, non-surgical delivery method to the brain, and an automated recovery plan if software updates fail. BCIs are designed to run for 12-15 years in implanted environments with no physical access. The onboard software is critical to processing signals collected from the brain. Delivering timely updates with the ability to roll back to stable versions if something goes wrong during the update process is critical.

| Manufacturers | Regulators |
|---|---|
| 1. Provide wireless or other non-surgical means of software update. | 1. Require that software updates have at least one non-surgical delivery option. |
| 2. State clear support windows for each update (e.g., when a patient needs to update their device next). | 2. Devices must check the integrity of provided updates before upgrading. |
| 3. Build integrity checks at the download, transfer, and installation phases of the update process. | |

## 6.2 Authentication and Authorization

BCIs require wireless connectivity to off-brain computers or other nodes within the brain to collect, transmit, process, and store data. Most regulators require "best effort" security but do not provide examples of strong authentication and authorization schemes. Device manufacturers should implement an authorization scheme for accessing BCI data and settings on the device. Ideally, this login process will implement separate and least privilege: a user can log in with only the permissions needed for the task at hand. An example would be two types of logins: a read-only login for viewing data, and a separate login for making changes to BCI settings. The patient or clinician may use the read-only login from various devices, and the edit login is only used when needed from trusted devices. The patient must retain the ability to reset or block a login.

| Manufacturers | Regulators |
|---|---|
| 1. Implement technical controls for authentication and authorization when accessing BCI settings or data | 1. Require authentication and authorization for accessing BCI settings or data |
| 2. Allow the patient to grant read-only or read-write access for logins | 2. Require use of least-privilege (ability to separate read and write access) |

3. Allow BCI patients to reset logins

3. Encourage further adoption of security best practices, like two-factor authentication

## 6.3 Minimized Attack Surface

BCIs are similar to devices in the Internet of Things: they are always-on and benefit from a constant network connection. While connectivity has its advantages, this same connection also exposes devices to cyber-attacks. BCIs only need a connection when transferring data or updating settings, which are both done at regular intervals in a controlled environment. Thus, manufacturers should implement a feature to allow patients to enable or disable wireless connection on the BCI. A patient could toggle the BCI internet connection via a companion device like a mobile phone (paired over local Bluetooth). Manufacturers should also consider extending additional control to the patient regarding when and how devices can connect with the BCI. Regulators should require such a feature to minimize the attack surface for BCIs.

| Manufacturers | Regulators |
|---|---|
| 1. Allow BCI patients to control when network connections are allowed<br>2. Allow BCI patients to remove or replace previously trusted devices | 1. Require connection control features in BCI devices |

## 6.4 Encryption

Most leading-edge BCIs do not implement encryption due to power limitations (Sriram, Karageorgos, et al., 2023). The FDA requires encryption for health data. Karageorgo's HALO paper proved that novel hardware architectures can allow encryption routines without exhausting power budgets. Because physical access is not a primary concern for BCIs, it would be most effective and reasonable to require encryption with data in transit only (being sent to or from a remote computer and the BCI) to minimize when additional power for encryption is needed. For leading-edge devices like SCALO, this would further include encryption for data transferred between the BCI's different wireless nodes.

| Manufacturers | Regulators |
|---|---|
| 1. Implement encryption routines for data leaving the brain | 1. Require encryption for data in transit<br>2. Require encryption optional for data stored on the brain if no physical ports exist |

## 6.5 Adversarial AI Defense

Adversarial machine learning attacks are the primary risk to BCI devices, until devices gain further capability to run AI models on-device. Upadhayay and Behzadan detail one such attack with intentionally modified stimuli a BCI patient experiences in their environment. We suggest the following to counter this and future AI-assisted attacks against BCI devices.

| Manufacturers | Regulators |
|---|---|
| 1. Adversarial training of AI/ML technologies to<br>2. Include Cognitive Status monitoring technology – stop decision-making if adversarial input is detected and refer to the patient for the next action<br>3. Enable the user to select which inputs should be allowed into the BCI. | 1. Require proof of training against adversarial AI attacks |

| Recommendation | Manufacturers | Regulators | Problem(s) addressed |
|---|---|---|---|
| Software Updates | 1. Provide wireless or other non-surgical means of software update.<br>2. State clear support windows for each update (e.g., when a patient needs to update their device next).<br>3. Build integrity checks at the download, transfer, and installation phases of the update process. | 1. Require that software updates have at least one non-surgical delivery option<br>2. Devices must check the integrity of provided updates before upgrading | Software update – invalid update<br><br>Software update – malicious update |
| Authentication and Authorization | 1. Implement technical controls for authentication and authorization when accessing BCI settings or data<br>2. Allow the patient to grant read-only or read-write access for logins<br>3. Allow BCI patients to reset logins | 1. Require authentication and authorization for accessing BCI settings or data<br>2. Require use of least-privilege (ability to separate read and write access)<br>3. Encourage further adoption of security best practices, like two-factor authentication | Onboard settings update<br><br>Unauthorized access |
| Attack Surface | 1. Allow BCI patients to control when network connections are allowed<br>2. Allow BCI patients to remove or replace previously trusted devices | 1. Require connection control features in BCI devices | Device exposure or compromise |
| Encryption | 1. Implement encryption routines for data leaving the brain | 1. Require encryption for data in transit<br>2. Require encryption optional for data stored on the brain if no physical ports exist | Local storage – unencrypted data<br><br>Data transmission |
| Adversarial AI Attacks | 1. Adversarial training of AI/ML technologies to | 1. Require proof of training against adversarial AI attacks | Adversarial AI attacks or use |

| | 2. Include Cognitive Status monitoring technology – stop decision-making if adversarial input is detected and refer to the patient for the next action 3. Enable the user to select which inputs should be allowed into the BCI. | | |
|---|---|---|---|

Table 3 - Recommendations Summary

## 7. Conclusion

BCIs present their own novel security issues that differ from standard issues faced by other FDA Class III medical devices. Real-time connections to BCIs provide efficient, non-surgical means to adjust devices to patient needs. However, they also pose a cybersecurity risk for an industry that is only starting to tackle cybersecurity challenges. We recommended and detailed four improvements that can set manufacturers and regulators on the path to more secure devices. If these recommendations are ignored, patient safety and privacy are at risk. As shown by Li, Class III devices are not immune to cyber-attacks (Li et al., 2011). Failure to implement over-the-air software updates means more patients will not regularly update devices when vulnerabilities are discovered. This leads to takeover risks and compromises patient data. Failure to implement strong login schemes will result in clinical personal health and genetic data being leaked or stolen. Failure to limit unnecessary network connections overly exposes BCIs to network-based cyber-attacks, increasing the risk of takeovers and data compromise with each attempt. High-risk devices should not be introduced to the public until essential security issues are resolved. Now is the time to implement the proper measures and prevent foreseeable problems.

**Acknowledgments**

We thank Abhishek Bhattacharjee, Raghav Pothukuchi, Muhammed Ugur, Michael J. Fischer, Jonathan Hochman, Tina Lu, Ted Wittenstein, and the Yale Schmidt Program on Artificial Intelligence, Emerging Technologies, and National Power for their helpful comments and discussions.

Thank you to our reviewers for their comments and insight on risks presented by AI to implanted BCIs. TS's work was supported by the Yale Digital Ethics Center Director's Fellowship and the Pauli Murray Mellon Research Fellowship for Seniors at Yale University.

**Declarations**

The authors declare that the research was conducted without any commercial or financial relationships that could be construed as a conflict of interest.

# References

Binnendijk, A., Marler, T., & Bartels, E. M. (2020). *Brain-Computer Interfaces: U.S. Military Applications and Implications, An Initial Assessment*. RAND Corporation. https://www.rand.org/pubs/research_reports/RR2996.html

Chase, M., Coley, S. C., Daldos, R., & Zuk, M. (2023). *Next Steps Toward Managing Legacy Medical Device Cybersecurity Risks*. https://www.mitre.org/news-insights/publication/next-steps-toward-managing-legacy-medical-device-cybersecurity-risks

Cloudflare. (n.d.). *What is lateral movement in cyber security?* Retrieved October 30, 2024, from https://www.cloudflare.com/learning/security/glossary/what-is-lateral-movement/

Denning, T., Matsuoka, Y., & Kohno, T. (2009). Neurosecurity: Security and privacy for neural devices. *Neurosurgical Focus*, *27*(1), E7. https://doi.org/10.3171/2009.4.FOCUS0985

FBI. (2022, September 12). *Unpatched and Outdated Medical Devices Provide Cyber Attack Opportunities*. American Heart Association. https://www.aha.org/system/files/media/file/2022/09/fbi-pin-tlp-white-unpatched-and-outdated-medical-devices-provide-cyber-attack-opportunities-sept-12-2022.pdf

FDA. (2020, March 3). *SweynTooth Cybersecurity Vulnerabilities May Affect Certain Medical Devices: FDA Safety Communication*. https://public4.pagefreezer.com/browse/FDA/08-02-2023T11:48/https://www.fda.gov/medical-devices/safety-communications/sweyntooth-cybersecurity-vulnerabilities-may-affect-certain-medical-devices-fda-safety-communication

Finlayson, S. G., Bowers, J. D., Ito, J., Zittrain, J. L., Beam, A. L., & Kohane, I. S. (2019). Adversarial attacks on medical machine learning. *Science*, *363*(6433), 1287–1289. https://doi.org/10.1126/science.aaw4399

FIRST. (2023, November 9). *CVSS v4.0 Specification Document*. FIRST — Forum of Incident Response and Security Teams. https://www.first.org/cvss/specification-document

Glannon, W. (2016). Ethical issues in neuroprosthetics. *Journal of Neural Engineering*, *13*(2), 021002. https://doi.org/10.1088/1741-2560/13/2/021002

Harris, R. (2020, February 17). *Hacking brain-computer interfaces*. ZDNET. https://www.zdnet.com/article/hacking-brain-computer-interfaces/

Health, C. for D. and R. (2023a). How to Determine if Your Product is a Medical Device. *FDA*. https://www.fda.gov/medical-devices/classify-your-medical-device/how-determine-if-your-product-medical-device

Health, C. for D. and R. (2023b). Your Clinical Decision Support Software: Is It a Medical Device? *FDA*. https://www.fda.gov/medical-devices/software-medical-device-samd/your-clinical-decision-support-software-it-medical-device

*Health Insurance Portability and Accountability Act of 1996 (HIPAA) | CDC*. (2022, June 28). https://www.cdc.gov/phlp/publications/topic/hipaa.html

*H.R.2617—117th Congress (2021-2022): Consolidated Appropriations Act, 2023*. (2021, November 3). https://www.congress.gov/bill/117th-congress/house-bill/2617/text

Ienca, M., Fins, J. J., Jox, R. J., Jotterand, F., Voeneky, S., Andorno, R., Ball, T., Castelluccia, C., Chavarriaga, R., Chneiweiss, H., Ferretti, A., Friedrich, O., Hurst, S., Merkel, G., Molnár-Gábor, F., Rickli, J.-M., Scheibner, J., Vayena, E., Yuste, R., & Kellmeyer, P. (2022). Towards a Governance Framework for Brain Data. *Neuroethics*, *15*(2), 20. https://doi.org/10.1007/s12152-022-09498-8

Ienca, M., & Haselager, P. (2016). Hacking the brain: Brain–computer interfacing technology and the ethics of neurosecurity. *Ethics and Information Technology*, *18*(2), 117–129. https://doi.org/10.1007/s10676-016-9398-9

Karageorgos, I., Sriram, K., Vesely, J., Wu, M., Powell, M., Borton, D., Manohar, R., & Bhattacharjee, A. (2020a). Hardware-Software Co-Design for Brain-Computer Interfaces.

*2020 ACM/IEEE 47th Annual International Symposium on Computer Architecture (ISCA)*, 391–404. https://doi.org/10.1109/ISCA45697.2020.00041

Karageorgos, I., Sriram, K., Vesely, J., Wu, M., Powell, M., Borton, D., Manohar, R., & Bhattacharjee, A. (2020b). Hardware-Software Co-Design for Brain-Computer Interfaces. *2020 ACM/IEEE 47th Annual International Symposium on Computer Architecture (ISCA)*, 391–404. https://doi.org/10.1109/ISCA45697.2020.00041

Lake, J. (2022, June 14). What is the CIA triad—Confidentiality, integrity and availability? *Comparitech*. https://www.comparitech.com/blog/information-security/confidentiality-integrity-availability/

Lebedev, M. A., & Nicolelis, M. A. L. (2006). Brain–machine interfaces: Past, present and future. *Trends in Neurosciences*, *29*(9), 536–546. https://doi.org/10.1016/j.tins.2006.07.004

Li, C., Raghunathan, A., & Jha, N. K. (2011). Hijacking an insulin pump: Security attacks and defenses for a diabetes therapy system. *2011 IEEE 13th International Conference on E-Health Networking, Applications and Services*, 150–156. https://doi.org/10.1109/HEALTH.2011.6026732

Liv, N. (2021). Neurolaw: Brain-Computer Interfaces. *University of St. Thomas Journal of Law & Public Policy*, *15*, 328–355.

Liv, N., & Greenbaum, D. (2023). Cyberneurosecurity. In V. Dubljević & A. Coin (Eds.), *Policy, Identity, and Neurotechnology: The Neuroethics of Brain-Computer Interfaces* (pp. 233–251). Springer International Publishing. https://doi.org/10.1007/978-3-031-26801-4_13

López Madejska, V. M., López Bernal, S., Martínez Pérez, G., & Huertas Celdrán, A. (2024). Impact of neural cyberattacks on a realistic neuronal topology from the primary visual cortex of mice. *Wireless Networks*, *30*(9), 7391–7405. https://doi.org/10.1007/s11276-023-03649-2

McGee, E. M. (2014). Brain–Computer Interfaces: Ethical and Policy Considerations. In *Implantable Bioelectronics* (pp. 411–433). John Wiley & Sons, Ltd. https://doi.org/10.1002/9783527673148.ch19

Ministry of Science and Technology (MOST; 科学技术部; 科技部). (2024). *Ethics Guidelines for Brain-Computer Interface Research* (B. Murphy, Trans.).

Newman, L. H. (2018, August 9). A New Pacemaker Hack Puts Malware Directly on the Device. *Wired.* https://www.wired.com/story/pacemaker-hack-malware-black-hat/

*Next-Generation Nonsurgical Neurotechnology.* (n.d.). Retrieved April 17, 2025, from https://www.darpa.mil/research/programs/next-generation-nonsurgical-neurotechnology

Pycroft, L., Boccard, S. G., Owen, S. L. F., Stein, J. F., Fitzgerald, J. J., Green, A. L., & Aziz, T. Z. (2016). Brainjacking: Implant Security Issues in Invasive Neuromodulation. *World Neurosurgery, 92*, 454–462. https://doi.org/10.1016/j.wneu.2016.05.010

QianQian Li, Ding Ding, & Conti, M. (2015). Brain-Computer Interface applications: Security and privacy challenges. *2015 IEEE Conference on Communications and Network Security (CNS)*, 663–666. https://doi.org/10.1109/CNS.2015.7346884

Regulation (EU) 2017/745 of the European Parliament and of the Council of 5 April 2017 on Medical Devices, Amending Directive 2001/83/EC, Regulation (EC) No 178/2002 and Regulation (EC) No 1223/2009 and Repealing Council Directives 90/385/EEC and 93/42/EEC (Text with EEA Relevance. ), 117 OJ L (2017). http://data.europa.eu/eli/reg/2017/745/oj/eng

Schwartz, N. (2024, October 2). *One Year Later: The Impact of the PATCH Act and Final Premarket Guidance on Medical Device Cybersecurity.* https://www.medcrypt.com/blog/one-year-later-the-impact-of-the-patch-act-and-final-premarket-guidance-on-medical-device-cybersecurity

Seh, A. H., Zarour, M., Alenezi, M., Sarkar, A. K., Agrawal, A., Kumar, R., & Ahmad Khan, R. (2020). Healthcare Data Breaches: Insights and Implications. *Healthcare*, *8*(2), Article 2. https://doi.org/10.3390/healthcare8020133

Shupe, L. E., Miles, F. P., Jones, G., Yun, R., Mishler, J., Rembado, I., Murphy, R. L., Perlmutter, S. I., & Fetz, E. E. (2021a). Neurochip3: An Autonomous Multichannel Bidirectional Brain-Computer Interface for Closed-Loop Activity-Dependent Stimulation. *Frontiers in Neuroscience*, *15*. https://doi.org/10.3389/fnins.2021.718465

Shupe, L. E., Miles, F. P., Jones, G., Yun, R., Mishler, J., Rembado, I., Murphy, R. L., Perlmutter, S. I., & Fetz, E. E. (2021b). Neurochip3: An Autonomous Multichannel Bidirectional Brain-Computer Interface for Closed-Loop Activity-Dependent Stimulation. *Frontiers in Neuroscience*, *15*, 718465. https://doi.org/10.3389/fnins.2021.718465

Sirbu, R. (2023). *Premarket Approval for AI/ML Software as a Medical Device (SaMD): The Past, Present, and Future of Cybersecurity in the U.S. Healthcare System.*

Sriram, K., Karageorgos, I., Wen, X., Vesely, J., Lindsay, N., Wu, M., Khazan, L., Pothukuchi, R. P., Manohar, R., & Bhattacharjee, A. (2023). HALO: A Hardware–Software Co-Designed Processor for Brain–Computer Interfaces. *IEEE Micro*, *43*(3), 64–72. https://doi.org/10.1109/MM.2023.3258907

Sriram, K., Pothukuchi, R. P., Gerasimiuk, M., Ugur, M., Ye, O., Manohar, R., Khandelwal, A., & Bhattacharjee, A. (2023). SCALO: An Accelerator-Rich Distributed System for Scalable Brain-Computer Interfacing. *Proceedings of the 50th Annual International Symposium on Computer Architecture*, 1–20. https://doi.org/10.1145/3579371.3589107

Steindl, E. (2024). Consumer neuro devices within EU product safety law: Are we prepared for big tech ante portas? *Computer Law & Security Review*, *52*, 105945. https://doi.org/10.1016/j.clsr.2024.105945

Thomas, M. A. (2022). Distinguishing Cyberattacks by Difficulty. *International Journal of Intelligence and CounterIntelligence*, *35*(4), 784–805. https://doi.org/10.1080/08850607.2021.2018565

Upadhayay, B., & Behzadan, V. (2023). Adversarial Stimuli: Attacking Brain-Computer Interfaces via Perturbed Sensory Events. *2023 IEEE International Conference on Systems, Man, and Cybernetics (SMC)*, 3061–3066. https://doi.org/10.1109/SMC53992.2023.10394505

*URGENT/11 Cybersecurity Vulnerabilities in a Widely-Used Third-Party Software Component May Introduce Risks During Use of Certain Medical Devices: FDA Safety Communication.* (2019, October 1). https://public4.pagefreezer.com/browse/FDA/16-06-2022T13:39/https://www.fda.gov/medical-devices/safety-communications/urgent11-cybersecurity-vulnerabilities-widely-used-third-party-software-component-may-introduce

*US Healthcare at risk: Strengthening resiliency against ransomware attacks.* (n.d.). Retrieved October 27, 2024, from https://www.microsoft.com/en-us/security/security-insider/emerging-threats/us-healthcare-at-risk-strengthening-resiliency-against-ransomware-attacks

Wilton, R. (2021, April 16). Encryption unlocks the benefits of a thriving, trustworthy Internet. *Internet Society*. https://www.internetsociety.org/resources/doc/2021/encryption-unlocks-the-benefits-of-a-thriving-trustworthy-internet/

Wolpaw, J., & Wolpaw, E. W. (Eds.). (2012). *Brain–Computer Interfaces: Principles and Practice*. Oxford University Press. https://doi.org/10.1093/acprof:oso/9780195388855.001.0001

Zettler, P. J., & Lietzan, E. (2021). Regulating Medical Devices in the United States. In D. Orentlicher & T. K. Hervey (Eds.), *The Oxford Handbook of Comparative Health Law* (p. 0). Oxford University Press. https://doi.org/10.1093/oxfordhb/9780190846756.013.59